%Paper: hep-ph/9212205
%From: lee@theory3.caltech.edu (C. L. Lee)
%Date: Tue, 1 Dec 92 15:42:55 PST

% DEFINITIONS

\input jytex.tex
\typesize = 12pt
\baselinestretch = 1200
%input figmac
\newcount\fignum  \fignum=0
\def\fignumstyle#1{\global\expandafter\let\expandafter
	\fignumtype\csname#1\endcsname}
\def\figlabel#1{\global\advance\fignum by1
  \label{#1}{\hbox{\fignumtype\fignum}}\putlab{#1}}
\def\Figure#1{\hbox{Fig.~\figlabel{#1}}}

\fignumstyle{arabic}

\newcount\tabnum  \tabnum=0
\def\tabnumstyle#1{\global\expandafter\let\expandafter
	\tabnumtype\csname#1\endcsname}
\def\tablabel#1{\global\advance\tabnum by1
  \label{#1}{\hbox{\tabnumtype\tabnum}}\putlab{#1}}

\tabnumstyle{arabic}
% input tables.tex
\newbox\hdbox%
\newcount\hdrows%
\newcount\multispancount%
\newcount\ncase%
\newcount\ncols% This is the number of primary text columns in the table.
\newcount\nrows%
\newcount\nspan%
\newcount\ntemp%
\newdimen\hdsize%
\newdimen\newhdsize%
\newdimen\parasize%
\newdimen\spreadwidth%
\newdimen\thicksize%
\newdimen\thinsize%
\newdimen\tablewidth%
\newif\ifcentertables%
\newif\ifendsize%
\newif\iffirstrow%
\newif\iftableinfo%
\newtoks\dbt%
\newtoks\hdtks%
\newtoks\savetks%
\newtoks\tableLETtokens%
\newtoks\tabletokens%
\newtoks\widthspec%
%
%  Book-keeping stuff--see how often these macros are called.
%
%  MOD RFC 900221.
%  Removed usage logging:  it's too complicated under VM/XA.
%\immediate\write15{%
%CP SMSG GJMSINK TEXTABLE --> TABLE MACROS V. 851121 JOB = \jobname%
%}%
%
%  Turn on table diagnostics.
%
\tableinfotrue%
\catcode`\@=11%  Allows use of "@" in macro names, like PLAIN.TEX does.
%  Debugging aid.  Writes #1 on the
%                                    user's terminal and in the log file.
%
%  Define the \tstrut height, depth in terms of the x_height parameter.
%
\def\tstrut{\vrule height3.1ex depth1.2ex width0pt}%
\def\and{\char`\&}%  Allows us to get an `&' in the text.  This is the
%                    same as using the PLAIN TeX macro \&.
\def\tablerule{\noalign{\hrule height\thinsize depth0pt}}%
\thicksize=1.5pt%  Default thickness for fat rules.  The user should feel
%                  free to change this to his preference.
\thinsize=0.6pt%   Default thickness for thin rules.
\def\thickrule{\noalign{\hrule height\thicksize depth0pt}}%
\def\ctr#1{\hfil\ #1\hfil}%
%
%
%
%  Here are things for controlling the width of the finished table.
%
\tablewidth=-\maxdimen%
\spreadwidth=-\maxdimen%
\def\tabskipglue{0pt plus 1fil minus 1fil}%
%
%  Stuff for centering or not.
%
\centertablestrue%
%
%
%
%  \vctr vertically centers its argument in the row.
%
\parasize=4in%
\gdef\ARGS{########}%  Produces the correct number of #'s in the preamble
%                      by the time eveything is expanded and \halign sees
%                      it.
\gdef\headerARGS{####}%  Same as \ARGS, but used in \header macros.
\def\@mpersand{&}%  Allows us to get alignment tab characters later
%                   when we have made the character "&" an active macro.
{\catcode`\|=13%  Make |'s locally active.
\gdef\letbarzero{\let|0}%  Globally define a macro that allows us to
%                          keep active |'s from being expanded in edef's.
\gdef\letbartab{\def|{&&}}%
\gdef\letvbbar{\let\vb|}%
%  This \def will cause active |'s read by
%                            \ruledtable to be converted into double
%                            alignment tabs.
}%  End of locally active |'s.
{\catcode`\&=4%  Make these alignment tabs.
\def\ampskip{&\omit\hfil&}%  This local macro skips a vertical rule.
\catcode`\&=13%  Now make &'s into active macros.
\let&0%  This allows us to expand \ampskip in the next \xdef without
%        attempting to expand the & and getting an "undefined control
%        sequence" error.
\xdef\letampskip{\def&{\ampskip}}%
\gdef\letnovbamp{\let\novb&\let\tab&}
%  This will cause active &'s read by
%                                   \ruledtable to be converted into
%                                   double tabs and an \omit'ted \vrule.
}%  End of locally active &'s.
\def\begintable{%  Here we make |'s and &'s active characters so we can
%                  interpret them as macros.  Note that this action is
%                  true only until we encounter the matching \endgroup
%                  token later at the end of the \ruledtable macro.
   \begingroup%
   \catcode`\|=13\letbartab\letvbbar%
   \catcode`\&=13\letampskip\letnovbamp%
   \def\multispan##1{%  We must redefine \multispan to count the number
%                       of primary columns, not physical columns.
      \omit \mscount##1%
      \multiply\mscount\tw@\advance\mscount\m@ne%
      \loop\ifnum\mscount>\@ne \sp@n\repeat%
   }%  End of \multispan macro.
   \def\|{%
      &\omit\widevline&%
   }%
   \ruledtable%  Now we call \ruledtable to do the real work.
}%  End of \begintable macro.
\long\def\ruledtable#1\endtable{%
%
%  This macro reads in the user's data entries
%  and converts them into a ruled table.
%
%  Important note:  Many macros and parameters are re-defined here, and
%  these must be kept local to the table macros to avoid conflict with
%  their use outside of tables.  This is done by the \begingroup token
%  macro \begintable and the \endgroup token at the end of
%  this macro.
%
   \offinterlineskip%  Needed to make rules touch each other.
   \tabskip 0pt%  Needed for same reason as \offinterlineskip.
   \def\widevline{\vrule width\thicksize}%  Make outer \vrule's wider.
   \def\endrow{\@mpersand\omit\hfil\crnorm\@mpersand}%
   \def\crthick{\@mpersand\crnorm\thickrule\@mpersand}%
   \def\crthickneg##1{\@mpersand\crnorm\thickrule
          \noalign{{\skip0=##1\vskip-\skip0}}\@mpersand}%
   \def\crnorule{\@mpersand\crnorm\@mpersand}%
   \def\crnoruleneg##1{\@mpersand\crnorm
          \noalign{{\skip0=##1\vskip-\skip0}}\@mpersand}%
   \let\nr=\crnorule%  A shorter abbreviation.
   \def\endtable{\@mpersand\crnorm\thickrule}%
   \let\crnorm=\cr%  Allows us to use \cr for our own purposes.
%
%  Cause user-typed \cr's to follow a row with a \tablerule.
%
   \edef\cr{\@mpersand\crnorm\tablerule\@mpersand}%
   \def\crneg##1{\@mpersand\crnorm\tablerule
          \noalign{{\skip0=##1\vskip-\skip0}}\@mpersand}%
   \let\ctneg=\crthickneg
   \let\nrneg=\crnoruleneg
   \the\tableLETtokens%  Get the user's extra \let's, if any.
%
%  Put the data entries into a token register so we can scan through them
%  and see what the user is asking us to do.
%
   \tabletokens={&#1}%  We add an extra alignment tab to the beginning
%                       of the first row to allow for the first \vrule.
%
%  Now count how many rows are in the table and return the result in
%  count register \nrows; do the same for columns, and return that
%  in register \ncols.
%
   \countROWS\tabletokens\into\nrows%
   \countCOLS\tabletokens\into\ncols%
%
%  Now do a little arithmetic to convert the number of primary columns
%  into the number of physical columns that the alignment preamble must
%  prepare for;  similarly for rows.
%
   \advance\ncols by -1%
   \divide\ncols by 2%
   \advance\nrows by 1%
%
%  Tell the user how many rows and columns we found in his data, if he
%  wants to know.
%
   \iftableinfo %
      \immediate\write16{[Nrows=\the\nrows, Ncols=\the\ncols]}%
   \fi%
%
%  Now we actually go ahead and produce the table.
%
   \ifcentertables
      \ifhmode \par\fi%  Make sure we are in vertical mode.
      \line{%  The final table comes out as an \hbox of width the \hsize.
      \hss%  The final table will be centered left-to-right.
   \else %
      \hbox{%
   \fi
      \vbox{%
         \makePREAMBLE{\the\ncols}%  Generate the preamble.
         \edef\next{\preamble}%  This line and the next line force the
         \let\preamble=\next%    expansion of all \ARGS tokens into the
%                                appropriate number of #'s.
         \makeTABLE{\preamble}{\tabletokens}%  Go do the \halign here.
      }%  End of \vbox.
      \ifcentertables \hss}\else }\fi%  Finish the centering effect.
%                                       It is important that no spaces
%                                       follow the two `}' here.
%  }%  End of \line.
   \endgroup%  Return all local macros and parameters to their outside
%              values.
   \tablewidth=-\maxdimen%  Reset \tablewidth to normal.
   \spreadwidth=-\maxdimen% Same for \spreadwidth.
}%  End of macro \ruledtable.
\def\makeTABLE#1#2{%  Does an \halign for the \ruledtable macro.
   {%  Start of local parameter values.
   \let\ifmath0%     These macros would cause trouble if they were to be
   \let\header0%     expanded in the following \xdef; we \let them be
   \let\multispan0%  equal to a digit, because digits can't be expanded.
%
%  Set up the width specification here.
%
   \ncase=0%
   \ifdim\tablewidth>-\maxdimen \ncase=1\fi%
   \ifdim\spreadwidth>-\maxdimen \ncase=2\fi%
   \relax%  This \relax is absolutely necessary, without it the following
%           \ifcase will always take \ncase=0.
%
   \ifcase\ncase %
      \widthspec={}%
   \or %
      \widthspec=\expandafter{\expandafter t\expandafter o%
                 \the\tablewidth}%
   \else %
      \widthspec=\expandafter{\expandafter s\expandafter p\expandafter r%
                 \expandafter e\expandafter a\expandafter d%
                 \the\spreadwidth}%
   \fi %
%\out{Widthspec=[\the\widthspec]}%
%\out{Preamble=[\preamble]}%
   \xdef\next{%  We must force the preamble to be expanded BEFORE the
      \halign\the\widthspec{%
%        \halign is done;  this \edef\next{...}\next construction
%                does the trick.
      #1%  This is the preamble text.
      \noalign{\hrule height\thicksize depth0pt}%  Makes the top \hrule.
      \the#2\endtable%  This is the main body.
%
%     \noalign{\hrule height0.7pt depth0pt}%  Makes the last \hrule.
      }%  End of \halign.
   }%  End of \next.
   }%  End of local values.
   \next%  This \next must be outside of the local values, because now
%          we want those troublesome macros in the \let's above to have
%          their normal actions.
}%  End of macro \makeTABLE.
\def\makePREAMBLE#1{%  This macro generates the necessary preamble for a
%                      ruled table with #1 primary columns.
%                      (Primary columns means the number of columns NOT
%                       counting those used for vertical rules.)
   \ncols=#1%  Get the number of columns desired.
   \begingroup%  Start local parameter definitions.
   \let\ARGS=0%  This is the key to the whole thing; it prevents \ARGS
%                from being expanded in the following \edef's.
   \edef\xtp{\widevline\ARGS\tabskip\tabskipglue%
   &\ctr{\ARGS}\tstrut}%  A 1-column preamble.  Gets the sizing right.
   \advance\ncols by -1%  One column has been generated; decrement the
%                         counter.
   \loop%  Append as many further columns as needed to the preamble.
      \ifnum\ncols>0 %
      \advance\ncols by -1%
      \edef\xtp{\xtp&\vrule width\thinsize\ARGS&\ctr{\ARGS}}%
   \repeat
   \xdef\preamble{\xtp&\widevline\ARGS\tabskip0pt%
   \crnorm}%  Adds the last \vrule.
   \endgroup%  End of local parameters.
}%  End of macro \makePREAMBLE.
\def\countROWS#1\into#2{%  This counts the number of rows in #1 by
%                          looking for control sequences that end a row,
%                          e.g., \cr, \crthick, etc., and puts the result
%                          into count register #2.
   \let\countREGISTER=#2%
   \countREGISTER=0%
%  \out{In countROWS:  tokens are [\the#1]}%
   \expandafter\ROWcount\the#1\endcount%
}%
\def\ROWcount{%
   \afterassignment\subROWcount\let\next= %
}%
\def\subROWcount{%
%  \out{In subROWcount:  next is [\meaning\next]}%  Debugging aid.
   \ifx\next\endcount %
      \let\next=\relax%
   \else%
      \ncase=0%
      \ifx\next\cr %
         \global\advance\countREGISTER by 1%
         \ncase=0%
      \fi%
      \ifx\next\endrow %
         \global\advance\countREGISTER by 1%
         \ncase=0%
      \fi%
      \ifx\next\crthick %
         \global\advance\countREGISTER by 1%
         \ncase=0%
      \fi%
      \ifx\next\crnorule %
         \global\advance\countREGISTER by 1%
         \ncase=0%
      \fi%
      \ifx\next\crthickneg %
         \global\advance\countREGISTER by 1%
         \ncase=0%
      \fi%
      \ifx\next\crnoruleneg %
         \global\advance\countREGISTER by 1%
         \ncase=0%
      \fi%
      \ifx\next\crneg %
         \global\advance\countREGISTER by 1%
         \ncase=0%
      \fi%
      \ifx\next\header %
%     \out{In subROWcount:  next=header, ncase set=1}%
         \ncase=1%
      \fi%
%     \out{In subROWcount:  ncase is [\the\ncase]}%
      \relax%
      \ifcase\ncase %
         \let\next\ROWcount%
%        \out{subROWcount---> ncase=\the\ncase}%
      \or %
         \let\next\argROWskip%
%        \out{subROWcount---> ncase=\the\ncase}%
      \else %
      \fi%
   \fi%
%  \out{subROWcount---> NEXT=\meaning\next}%
   \next%
}%  End of macro \subROWcount.
\def\counthdROWS#1\into#2{%
\dvr{10}%
   \let\countREGISTER=#2%
   \countREGISTER=0%
\dvr{11}%
%  \out{In counthdROWS:  tokens are [\the#1]}%
\dvr{13}%
   \expandafter\hdROWcount\the#1\endcount%
\dvr{12}%
}%
\def\hdROWcount{%
   \afterassignment\subhdROWcount\let\next= %
}%
\def\subhdROWcount{%
%\out{In subhdROWcount:  next is [\meaning\next]}%
   \ifx\next\endcount %
      \let\next=\relax%
   \else%
      \ncase=0%
      \ifx\next\cr %
         \global\advance\countREGISTER by 1%
         \ncase=0%
      \fi%
      \ifx\next\endrow %
         \global\advance\countREGISTER by 1%
         \ncase=0%
      \fi%
      \ifx\next\crthick %
         \global\advance\countREGISTER by 1%
         \ncase=0%
      \fi%
      \ifx\next\crnorule %
         \global\advance\countREGISTER by 1%
         \ncase=0%
      \fi%
      \ifx\next\header %
%\out{In subhdROWcount:  next=header, ncase set=1}%
         \ncase=1%
      \fi%
%\out{In subhdROWcount:  ncase is [\the\ncase]}%
\relax%
      \ifcase\ncase %
         \let\next\hdROWcount%
%\out{subhdROWcount---> ncase=\the\ncase}%
      \or%
         \let\next\arghdROWskip%
%\out{subhdROWcount---> ncase=\the\ncase}%
      \else %
      \fi%
   \fi%
%\out{subhdROWcount---> NEXT=\meaning\next}%
   \next%
}%
{\catcode`\|=13\letbartab
\gdef\countCOLS#1\into#2{%
%  \out{In countCOLS:  tokens are [\the#1]}
   \let\countREGISTER=#2%
   \global\countREGISTER=0%
   \global\multispancount=0%
   \global\firstrowtrue
   \expandafter\COLcount\the#1\endcount%
   \global\advance\countREGISTER by 3%
   \global\advance\countREGISTER by -\multispancount
%  \out{countCOLS-->[\the\countREGISTER]}
}%
\gdef\COLcount{%
   \afterassignment\subCOLcount\let\next= %
}%
{\catcode`\&=13%
\gdef\subCOLcount{%
%\out{In subCOLcount: next is [\meaning\next]}
   \ifx\next\endcount %
      \let\next=\relax%
   \else%
      \ncase=0%
      \iffirstrow
         \ifx\next& %
            \global\advance\countREGISTER by 2%
            \ncase=0%
         \fi%
         \ifx\next\span %
            \global\advance\countREGISTER by 1%
            \ncase=0%
         \fi%
         \ifx\next| %
            \global\advance\countREGISTER by 2%
            \ncase=0%
         \fi
         \ifx\next\|
            \global\advance\countREGISTER by 2%
            \ncase=0%
         \fi
         \ifx\next\multispan
            \ncase=1%
            \global\advance\multispancount by 1%
         \fi
         \ifx\next\header
            \ncase=2%
         \fi
         \ifx\next\cr       \global\firstrowfalse \fi
         \ifx\next\endrow   \global\firstrowfalse \fi
         \ifx\next\crthick  \global\firstrowfalse \fi
         \ifx\next\crnorule \global\firstrowfalse \fi
         \ifx\next\crnoruleneg \global\firstrowfalse \fi
         \ifx\next\crthickneg  \global\firstrowfalse \fi
         \ifx\next\crneg       \global\firstrowfalse \fi
      \fi%  End of \iffirstrow.
\relax%\out{subCOL-->  ncase=[\the\ncase]}
% \out{subCOL-->  next=\meaning\next}
      \ifcase\ncase %
         \let\next\COLcount%
      \or %
         \let\next\spancount%
      \or %
         \let\next\argCOLskip%
      \else %
      \fi %
   \fi%
%  \out{subCOL-->  countREGISTER=[\the\countREGISTER]}
   \next%
}%
\gdef\argROWskip#1{%
%  Deletes the next balanced, undelimited argument from a
%                 token list.
% \out{---> Entering argROWskip <---}
% \out{In argROWskip:  deleted arg is [#1]}%
   \let\next\ROWcount \next%
}%  End of macro \argskip.
\gdef\arghdROWskip#1{%
%  Deletes the next balanced, undelimited argument from a
%                 token list.
% \out{---> Entering arghdROWskip <---}
% \out{In arghdROWskip:  deleted arg is [#1]}%
   \let\next\ROWcount \next%
}%  End of macro \arghdROWskip.
\gdef\argCOLskip#1{%
%  Deletes the next balanced, undelimited argument from a
%                 token list.
% \out{---> Entering argCOLskip <---}
% \out{In argCOLskip:  deleted arg is [#1]}%
   \let\next\COLcount \next%
}%  End of macro \argskip.
}%  End of active &'s.
}%  End of active |'s.
\def\spancount#1{%\out{spancount--->\meaning#1}
   \nspan=#1\multiply\nspan by 2\advance\nspan by -1%
   \global\advance \countREGISTER by \nspan
%  \out{number spancount--->\the\nspan; \the\countREGISTER}
   \let\next\COLcount \next}%
\def\dvr#1{\relax}%
% \omit\hfil%
% \parindent=0pt\hsize=1.1in\valign{%
% \vfil#\vfil&\vfil#\vfil\cr\hfil\hbox{\ Added to\ }\hfil&%
% \hfil\hbox{\ empty events\ }\hfil\cr}\hfil%
\def\header#1{%
\dvr{1}{\let\cr=\@mpersand%
\hdtks={#1}%
%\out{In header:  hdtks=[\the\hdtks]}%
\counthdROWS\hdtks\into\hdrows%
\advance\hdrows by 1%
\ifnum\hdrows=0 \hdrows=1 \fi%
%\out{In header:  Nhdrows=[\the\hdrows]}%
\dvr{5}\makehdPREAMBLE{\the\hdrows}%
%\out{In header:  headerpreamble=[\headerpreamble]}%
\dvr{6}\getHDdimen{#1}%
%\out{In header:  hdsize=[\the\hdsize]}%
%\striplastCR{#1}%
{\parindent=0pt\hsize=\hdsize{\let\ifmath0%
\xdef\next{\valign{\headerpreamble #1\crnorm}}}\dvr{7}\next\dvr{8}%
}%
}\dvr{2}}%  End of macro \header.
\def\makehdPREAMBLE#1{%This macro generates the necessary preamble for a
\dvr{3}%
%                      ruled table with \ncols primary columns.
%                      (Primary columns means the number of columns NOT
%                       counting those used for vertical rules.
\hdrows=#1%  Get the number of columns desired.
{%  Start local parameter definitions.
\let\headerARGS=0%
%  This is the key to the whole thing; it prevents \ARGS
\let\cr=\crnorm%
%                from being expanded in the followin \edef's.
\edef\xtp{\vfil\hfil\hbox{\headerARGS}\hfil\vfil}%
\advance\hdrows by -1%  One row has been generated; decrement the
%                         counter.
\loop%  Append as many further rows as needed to the preamble.
\ifnum\hdrows>0%
\advance\hdrows by -1%
\edef\xtp{\xtp&\vfil\hfil\hbox{\headerARGS}\hfil\vfil}%
\repeat%
\xdef\headerpreamble{\xtp\crcr}%
}%  End of local parameters.
\dvr{4}}%  End of \makehdPREAMBLE.
\def\getHDdimen#1{%
%\out{In getHDdimen:  Arg 1=[#1]}%
\hdsize=0pt%
\getsize#1\cr\end\cr%
}%  End of macro getHDdimen.
\def\getsize#1\cr{%
%\out{In getsize:  Arg 1=[#1]}%
%  Here we have to check arg#1 and see if the first token in #1 is an
%    \end; if so, we stop, else we check the width of arg#1.
%  We recall that each arg#1 will be terminated with a \cr token.
\endsizefalse\savetks={#1}%
%\out{In getsize:  the savetks = [\the\savetks]}%
\expandafter\lookend\the\savetks\cr%
%\out{In getsize:  ifendsize = [\meaning\ifendsize]}%
\relax \ifendsize \let\next\relax \else%
\setbox\hdbox=\hbox{#1}\newhdsize=1.0\wd\hdbox%
\ifdim\newhdsize>\hdsize \hdsize=\newhdsize \fi%
%\out{In getsize:  hdsize=[\the\hdsize]}%
%\out{In getsize:  newhdsize=[\the\newhdsize]}%
\let\next\getsize \fi%
\next%
}%
\def\lookend{\afterassignment\sublookend\let\looknext= }%
\def\sublookend{\relax%
%\out{In sublookend:  looknext = [\looknext]}%
\ifx\looknext\cr %
%\out{In sublooknext:  looknext=cr}%
\let\looknext\relax \else %
%\out{In sublooknext:  looknext/=cr}%
   \relax
   \ifx\looknext\end \global\endsizetrue \fi%
   \let\looknext=\lookend%
    \fi \looknext%
}%
%
%  Allow the user to make his own names for crthick, etc.
%
\def\tablelet#1{%
   \tableLETtokens=\expandafter{\the\tableLETtokens #1}%
}%
\catcode`\@=12%  Change @'s back to their normal category code.

\def\sec#1{{\bf\section{#1}}}   \sectionnumstyle{arabic}
\def\subsec#1{\medskip{\bf\subsection{#1}}}
\def\eq#1{\eqno\eqnlabel{#1}$$}
\def\puteq#1{eq.~(\puteqn{#1})}
\def\ref#1{\markup{[\putref{#1}]}}
\def\frac#1#2{{#1 \over #2}}

\def\Tr{{\rm Tr}}
\def\Fslash#1#2{#1\hskip -#2pt /}
\def\bra#1{\langle{#1}|}
\def\ket#1{|{#1}\rangle}
\def\cthH{\cos\theta_H}
\def\mDs{m_{D^*}}
\def\msp#1{\hskip#1em\relax}
\def\Dprop1{v^\prime \cdot p_{\pi}}
\def\Dprop2{v^\prime \cdot p_{\pi} + \Delta_D}
\def\Bprop{v \cdot p_{\pi} + \Delta_B}
\def\Gt#1{\tilde G_{#1,l}}
\def\Gh{\hat G}
\def\journal#1#2{{\sl #1} {\bf #2}}

% TITLE PAGE

\pagenumstyle{blank}
\hbox to\hsize{\footnotesize\baselineskip=12pt
  \hfil\vtop{\hbox{\strut CALT-68-1841} \hbox{\strut DOE RESEARCH
  AND} \hbox{\strut DEVELOPMENT REPORT}}}
\vskip1.in
\centertext{\baselinestretch=1000\bigsize\bf
An Analysis of the Decay $B \rightarrow D^* X \ell \bar\nu_\ell$ \\
with Predictions from Heavy Quark and Chiral Symmetry%
\footnote{This work was supported in part by the
  U.S. Department of Energy under Contract No. DE-AC0381-ER40050.}}
\vskip .25in \footnotenumstyle{arabic}
\centertext{Clarence L. Y. Lee%
\footnote[1]{Email: \tt lee@theory3.caltech.edu}}
\vskip .25in \footnotenumstyle{symbols}
\centertext{\it California Institute of Technology\\Pasadena, CA 91125}

\vskip .7in
\centerline{\bf ABSTRACT}
\medskip {\narrower \baselinestretch=1000
	This paper considers the implications of the heavy quark and
chiral symmetries for the semi-leptonic decay
$B \rightarrow D^* X \ell \bar \nu_\ell$. The general kinematic analysis
for decays of the form
{\sl pseudoscalar meson $\rightarrow$ vector meson $+$
   pseudoscalar meson $+$ lepton $+$ anti-lepton}
is presented. This formalism is applied to the above exclusive decay
which allows the differential decay rate to be expressed in a form
that is ideally suited for the experimental determination of the
different form factors for the process through angular distribution
measurements. Heavy quark and chiral symmetry predictions for the
form factors are presented, and the differential decay rate is
calculated in the kinematic region where chiral perturbation theory is
valid.
\par}

\vfill\lefttext{October 1992}\medskip
\newpage \pagenum=0 \pagenumstyle{arabic}

% BODY

\sec{INTRODUCTION}

	When the mass of a quark is taken to infinity with its
four-velocity held fixed, its strong interactions become independent of
its mass and spin, and depend only on its velocity. This gives rise to a
new SU($2N$) spin-flavor symmetry, for $N$ heavy quark flavors, that is
not manifest in the full theory of QCD. These symmetries are made
explicit in a heavy quark effective field theory (HQEFT) which has been
a powerful tool in understanding the strong dynamics of hadrons
containing a heavy quark.\ref{Wis91}

	Another symmetry of the strong interactions that has been known
for some time is that in the limit where the mass of the light $u,d,s$
quarks become massless, QCD has a $\rm SU(3)_L \times SU(3)_R$
chiral symmetry. This symmetry is spontaneously broken down to the
$\rm SU(3)_V$ vector subgroup by the strong interactions, and the
associated pseudo-Goldstone bosons are the light pseudoscalar octet
mesons $\pi,K,\eta$. Chiral Lagrangians which incorporate
this symmetry have been used extensively to study low energy
interactions involving the pseudoscalar mesons.

	Recently, a synthesis of the above heavy quark and chiral
symmetries has been achieved. This theory describes the low energy
interactions of hadrons containing a single heavy quark (which will
hereafter be referred to as a heavy hadron and excludes $Q \bar Q$
quarkonium bound states of heavy quarks) with the pseudoscalar octet
mesons.
\markup{[\putref{Wis92,Bur92,Yan92,Cho92,Lee92}]}
Reference [\putref{Lee92}] examined the constraints that these
symmetries place on $B_{\ell 4}$ and $D_{\ell 4}$ decays
(without a final state vector meson).
In this paper, similar considerations are applied to the decay
$B \rightarrow D^* X \ell \bar \nu_\ell$, where $X$ is a
pseudo-Goldstone boson, and the development here
parallels the one in that paper.

	The heavy meson chiral Lagrangian density which describes
the low momentum interactions of the ground state heavy mesons
with the pseudo-Goldstone bosons is given by\ref{Wis92}
$$\eqalignno{
 {\cal L} =& {f_{\pi}^2 \over 8}
  \Tr(\partial^{\mu} \Sigma \partial_{\mu} \Sigma^{\dagger})
   + \lambda_0 \Tr(m_q \Sigma + m_q \Sigma^{\dagger})
   -i\Tr \bar H_a v_{\mu} \partial^{\mu} H_a \cr
  &+ {i \over 2}
   \Tr \bar H_a H_b v_{\mu} (\xi^{\dagger} \partial^{\mu} \xi
       + \xi \partial^{\mu} \xi^{\dagger})_{ba}
   +{ig \over 2}\Tr \bar H_a H_b \gamma_{\mu} \gamma_5
   (\xi^{\dagger} \partial^{\mu} \xi
   - \xi \partial^{\mu} \xi^{\dagger})_{ba} \cr
  &+ \lambda_1 \Tr \bar H_a H_b
   (\xi m_q \xi + \xi^{\dagger} m_q \xi^{\dagger})_{ba}
  + \lambda_1^{\prime} \Tr \bar H_a H_a
   (m_q \Sigma + m_q \Sigma^{\dagger})_{bb} \cr
  &+ {\lambda_2 \over m_Q} \Tr \bar H_a \sigma_{\mu\nu} H_a
    \sigma^{\mu\nu} + \ldots, & \eqnlabel{L_HMCS} \cr}$$
where the ellipsis denotes terms containing more derivatives, factors of
the light quark mass matrix
$$ m_q = \pmatrix{m_u& 0& 0\cr 0& m_d& 0\cr 0& 0& m_s\cr} \eq{m_q}
which explicitly violates $\rm SU(3)_L \times SU(3)_R$ chiral
symmetry, or factors of $1/m_Q$ (where $m_Q$ is the mass of the heavy
quark of flavor $Q$) associated with the breaking of the
SU(2N) heavy quark spin-flavor symmetry. The light quark flavor
indices $a,b$ run over 1,2,3 (corresponding to $u,d,s$) and repeated
indices are implicitly summed.

	The heavy pseudoscalar and vector meson fields,
$P_a$ and $P_a^*$, are degenerate in the heavy quark mass limit, and
in the case where $Q$ is the $c$ quark,
$$(P_1,P_2,P_3) = (D^0,D^+,D_s)$$
and $$(P_1^*,P_2^*,P_3^*) = (D^{*0},D^{*+},D_s^*).$$
In the above Lagrangian, these heavy meson fields are
combined into the $4 \times 4$ matrix
$$ H_a = {1 + \Fslash{v}{6} \over 2}
         (P_{a\mu}^* \gamma^\mu - P_a \gamma_5), \eq{H_a}
and
$$ \bar H_a = \gamma^0 H_a^{\dagger} \gamma^0. \eq{Hbar_a}
The field $H_a$ is a doublet under the heavy quark spin symmetry
${\rm SU(2)}_v$ and an anti-triplet under chiral $SU(3)_V$:
$$ H_a \rightarrow S(H U^{\dagger})_a, \eq{Htransf}
where $S \in {\rm SU(2)}_v$, which is the symmetry group for a
single flavor of heavy quark at velocity $v$, and $U$ is defined below.

	The pseudo-Goldstone boson fields are incorporated into the
Lagrangian of \puteq{L_HMCS} via
$$\Sigma = \xi^2 \eq{Sigma} where
$$ \xi = \exp(iM/f_{\pi}) \eq{xi}
with
$$ M = \pmatrix{{1 \over \sqrt2}\pi^0 + {1 \over \sqrt6}\eta &
                 \pi^+ & K^+ \cr
                \pi^- & -{1 \over \sqrt2}\pi^0 +{1 \over \sqrt6}\eta &
                 K^0 \cr
                K^- & \bar K^0 & -\sqrt{2 \over 3} \eta \cr} \eq{M}
and $f_\pi \approx 132\, \rm MeV$ is the pion decay constant.
Under chiral $\rm SU(3)_L \times SU(3)_R$ transformations
\subequationnumstyle{alphabetic}
$$ \Sigma \rightarrow L \Sigma R^{\dagger} \eq{SigmaCtranf}
and
$$ \xi \rightarrow L \xi U^{\dagger} = U \xi R^{\dagger},
 \eq{xiCtranf}
\subequationnumstyle{blank}
where $L \in \rm SU(3)_L$, $R \in \rm SU(3)_R$, and $U$ is a
unitary matrix which depends on $L, R$ and is a function of
space-time through its non-linear dependence on the pseudo-Goldstone
boson fields.

	Equation (\puteqn{L_HMCS}) is the most general Lagrangian
invariant under both the heavy quark and chiral symmetries to
first order in the Goldstone boson momenta, $m_q$ and $1/m_Q$.
It is remarkable that these symmetries combine to constrain the
the Lagrangian so that at leading order there is only one
unknown coupling $g$ independent of the heavy quark flavor and
spin. This same coupling enters into the decay width for
$D^* \rightarrow D \pi$ (where $Q=c$):
$$\Gamma(D^{*+} \rightarrow D^0 \pi^+) = {g^2 \over 6 \pi}
 {|{\bf p}_\pi|^3 \over f_\pi^2} \eq{Dswidth}
A recent experimental measurement of this width gave the limit
$\Gamma(D^{*+} \rightarrow D^0 \pi^+) < 72 \; \rm keV$\ref{ACC92}
which translates to $g^2 < 0.4$.
There is no phase space for the corresponding decay when $Q=b$,
$B^* \rightarrow B \pi$. However, the exclusive semi-leptonic decay
$B \rightarrow D^* X \ell \bar \nu_\ell$ could be used to probe the
heavy flavor dependence of $g$.

	The kinematical formalism for treating an
exclusive decay of the form under consideration is developed in
Sect.2. In Sect. 3, the heavy quark and chiral symmetry predictions
for the form factors that appear in
$B \rightarrow D^* X \ell \bar \nu_\ell$ decay are calculated and
used to determine the rate for this process. Concluding remarks are
made in Sect.4.

\sec{KINEMATICAL ANALYSIS}

	In this section, the general kinematic analysis for
decays of the form
$$\hbox{pseudoscalar meson $\rightarrow$ vector meson +
   pseudoscalar meson + lepton + anti-lepton} \eq{p-vpln}
is presented. For definiteness, we consider the decay
$B \rightarrow D^* X \ell \bar \nu_\ell$; however, this formalism
is more generally applicable to any decay of the form given by
\puteq{p-vpln}. If $p_B,p_{D^*},p_X,p_\ell,p_{\bar\nu}$ are the
four-momenta of the $B,D^*\hbox{(which also has a polarization
vector $\varepsilon$)},X,\ell,\bar\nu_\ell$, respectively, then the
kinematics of the decay can be more conveniently
expressed in terms of quantities involving the following
combinations of these four-momenta.
\begin{eqnseries}[defPQLN]
$$\eqalignno{
 P &= p_{D^*} + p_X , &\eqnlabel{P}\cr
 Q &= p_{D^*} - p_X  &\eqnlabel{Q}\cr
 L &= p_\ell + p_{\bar \nu}  &\eqnlabel{L}\cr
 N &= p_\ell - p_{\bar \nu}  &\eqnlabel{N}\cr}$$
\end{eqnseries}
Apart from spin, four-body decay is kinematically parameterized by
five variables. By choosing these variables appropriately, one can
express the distribution for the decay in a form where the dependence
of the angular distribution on the hadronic and leptonic currents
factorizes. This can be achieved by the choice\ref{Cab65}
\item{i.} $s_H=P^2$, the effective mass of the hadron pair,
          $D^*$ and $X$;
\item{ii.} $s_L=L^2$, the effective mass of the lepton pair,
           $\ell$ and $\bar \nu_\ell$;
\item{iii.} $\theta_H$, the angle between the $D^*$ three-momentum
            in the $D^* X$ rest frame and the line of flight of the
            $D^* X$ in the rest frame of the $B$;
\item{iv.} $\theta_L$, the angle between the $\ell$ three-momentum
           in the $\ell \bar \nu_\ell$ rest frame and the line of
           flight of the
           $\ell \bar \nu_l$ in the rest frame of the $B$;
\item{v.} $\phi$, the angle from the normal of the plane formed
          by the hadron pair to the normal of the plane formed
          by the lepton pair.
\medskip

	In the following analysis, one finds that over much of
the available phase space including the region where
chiral perturbation theory is valid, terms that depend on the
mass of the lepton are suppressed, $m_\ell /s_L \ll 1$, so that
the lepton mass may be neglected. With $m_\ell = 0$,
\begin{eqnseries}[KinInvar]
$$\eqalignno{
 Q^2 &= 2(m_{D^*}{}^2 + m_X{}^2) - s_H = (\chi^2 - U^2) s_H,
   & \eqnlabel{Q^2} \cr
 N^2 &= - s_L, & \eqnlabel{N^2} \cr
 P \cdot L &= V = {m_B{}^2 - s_H - s_L \over 2}, & \eqnlabel{V} \cr
 P \cdot Q &= m_{D^*}{}^2 - m_X{}^2 = \chi s_H, & \eqnlabel{PQ} \cr
 P \cdot N &= W \cos\theta_L, &\eqnlabel{PN} \cr
 L \cdot N &= 0, &\eqnlabel{LN} \cr
 Q \cdot L &= \chi V + U W \cos\theta_H, &\eqnlabel{QL}\cr
 Q \cdot N &= (\chi W + U V \cos\theta_H)\cos\theta_L
  - U \sqrt{s_H s_L} \sin\theta_H \sin\theta_L \cos\phi, &\eqnlabel{QN}\cr
 \epsilon_{\mu\nu\rho\sigma} P^{\mu} Q^{\nu} L^{\rho} N^{\sigma} &=
   - U W \sqrt{s_H s_L} \sin\theta_H \sin\theta_L \sin\phi,
&\eqnlabel{PQLN}\cr}$$
\end{eqnseries}
In eqs.(\puteqn{KinInvar})
\subequationnumstyle{alphabetic}
$$ \chi = {m_{D^*}{}^2 - m_X{}^2 \over s_H}, \eq{chi}
$U$ is the magnitude of the $D^*$ three-momentum in the $D^* X$
rest frame,
$$ U = (s_H{}^2 +m_{D^*}{}^4 +m_X{}^4 -2 s_H m_{D^*}{}^2
        -2 s_H m_X{}^2 -2 m_{D^*}{}^2 m_X{}^2)^{1/2} /s_H, \eq{U}
and
$$ W = (V^2 - s_H s_L)^{1/2}. \eq{W}
\subequationnumstyle{blank}

	The invariant transition amplitude for the decay
$B \rightarrow D^* X \ell \bar \nu_\ell$ is given by
$$ {\cal M}_{fi} = {G_F \over \sqrt2} V_{cb}^*
        \bra{X(p_X) D^*(p_{D^*},\varepsilon)} \bar c \gamma_{\mu}
        (1-\gamma_5) b \ket{B(p_B)} \;
        \bar u(p_\ell) \gamma^{\mu} (1-\gamma_5) v(p_{\bar \nu}),
\eq{Mfi}
where $G_F$ is the Fermi constant and $V_{cb}^*$ is the
Cabibbo-Kobayashi-Maskawa matrix element for $b \rightarrow c$
transitions. The hadronic matrix element can be expressed in terms
of fifteen form factors:
$$\eqalignno{
 \bra{X(p_X) D^*(p_{D^*},\varepsilon)} & \bar c \gamma_{\mu}
  (1-\gamma_5) b \ket{B(p_B)} = \cr
  &\Bigl[i(a_+ \>\varepsilon^* \cdot P + b_+ \>\varepsilon^* \cdot p_B)
    + {w_+ \over 2} \epsilon_{\alpha\beta\gamma\delta}
      L^{\alpha} P^{\beta} Q^{\gamma} \varepsilon^{*\delta}\Bigr]P_{\mu}\cr
 +&\Bigl[i(a_- \>\varepsilon^* \cdot P + b_- \>\varepsilon^* \cdot p_B)
    + {w_- \over 2} \epsilon_{\alpha\beta\gamma\delta}
      L^{\alpha} P^{\beta} Q^{\gamma} \varepsilon^{*\delta}\Bigr]Q_{\mu}\cr
 +&\Bigl[i(c \>\varepsilon^* \cdot P + d \>\varepsilon^* \cdot p_B)
    + {w \over 2} \epsilon_{\alpha\beta\gamma\delta}
      L^{\alpha} P^{\beta} Q^{\gamma} \varepsilon^{*\delta}\Bigr]L_{\mu}\cr
 +&\> i f \varepsilon^*{}_{\mu} \cr
 +&\> g_+ \,\epsilon_{\mu\alpha\beta\gamma} L^{\alpha} P^{\beta}
   \varepsilon^{*\gamma}
 +g_- \,\epsilon_{\mu\alpha\beta\gamma} L^{\alpha} Q^{\beta}
   \varepsilon^{*\gamma}
 +r \epsilon_{\mu\alpha\beta\gamma} P^{\alpha} Q^{\beta}
   \varepsilon^{*\gamma} \cr
 +&\>(u_1 \>\varepsilon^* \cdot P + u_2 \>\varepsilon^* \cdot p_B)
    \epsilon_{\mu\alpha\beta\gamma}
    L^{\alpha} P^{\beta} Q^{\gamma}, &\eqnlabel{ffdecomp}\cr}$$
where the form factors $a_{\pm},b_{\pm},c,d,f,g_{\pm},r,u_1,u_2,
w,$ and $w_{\pm}$ are functions of $s_H,s_L$, and $\theta_H$.
The absolute value of the transition amplitude squared when
summed over the vector meson and lepton polarizations is then
$$\sum_{\rm spins} \bigl| {\cal M}_{fi} \bigr|^2 =
  {G_F{}^2 \over 2} |V_{cb}|^2 H_{\mu\nu} L^{\mu\nu},
\eq{aveM^2}
where
\subequationnumstyle{alphabetic}
$$\eqalignno{
 H_{\mu\nu} = \;
  &\bra{X(p_X) D^*(p_{D^*},\varepsilon)}
   \bar c \gamma_{\mu} (1-\gamma_5) b \ket{B(p_B)} \cr
  &\times \bra{X(p_X) D^*(p_{D^*},\varepsilon)}
   \bar c \gamma_{\nu} (1-\gamma_5) b \ket{B(p_B)}^*,
  &\eqnlabel{Hmunu}\cr
 L^{\mu\nu} = \;&4(L^{\mu} L^{\nu} -N^{\mu} N^{\nu} -s_L g^{\mu\nu}
   -i \epsilon^{\alpha\mu\beta\nu} L_{\alpha} N_{\beta}).
  &\eqnlabel{Lmunu}\cr}$$
\subequationnumstyle{blank}
Using eqs.(\puteqn{KinInvar}a--i), the differential decay rate
can then be written in the form
$$d^5 \Gamma =
  {G_F{}^2 \bigl| V_{cb} \bigr|^2 \over (4\pi)^6 \, m_{D^*}{}^3}
  U W I(s_H,s_L,\theta_H,\theta_L,\phi) ds_H \, ds_L \,
  d\cos\theta_H \, d\cos\theta_L \, d\phi, \eq{d5Gamma}
with
$$\eqalignno{I =  \; &I_1 + I_2 \cos2\theta_L
  + I_3 \sin^2\theta_L \cos2\phi + I_4 \sin2\theta_L \cos\phi
  + I_5 \sin\theta_L \cos\phi \cr
 &+ I_6 \cos\theta_L + I_7 \sin\theta_L \sin\phi
  + I_8 \sin2\theta_L \sin\phi + I_9 \sin^2\theta_L \sin2\phi
  & \eqnlabel{I} \cr}$$
where $I_j$, $1 \leq j \leq 9$, are functions of $s_H,s_L,
\theta_H$ only. As was alluded to earlier, the separation of the
variables $s_H,s_L,\theta_H$ from $\theta_L,\phi$ in
\puteq{I} is a direct consequence this particular choice
for the five variables parameterizing four-body decay.
The distribution functions $I_j$ can be written in a
compact form by introducing the following combinations of
kinematic factors and form factors.
\begin{eqnseries}[G]
$$\eqalignno{
 G_1 = & {1 \over 2 m_{D^*}}
  \biggl\{\lambda s_H [W a_+ +(\chi W +U V \cos\theta_H)a_-] \cr
    &\msp{3} +\biggl({m_B{}^2 +s_H -s_L \over 2}\lambda
                     +U W \cos\theta_H \biggr)
              [W b_+ +(\chi W +U V \cos\theta_H)b_-] \cr
    &\msp{3} +(\lambda W +U V \cos\theta_H)f \biggr\} &\eqnlabel{G1}\cr
 G_2 =& {U \sqrt{s_H s_L} \over 2 m_{D^*}}
  \biggl[(\lambda s_H) a_-
   +\biggl({m_B{}^2 +s_H -s_L \over 2}\lambda +U W \cos\theta_H \biggr)b_-
   + f \biggr] & \eqnlabel{G2} \cr
 G_3 =& \sqrt{s_H}
         \biggl\{[W a_+ +(\chi W +U V \cos\theta_H)a_-] \cr
           &\msp{2}+{m_B{}^2 +s_H -s_L \over 2 s_H}
            [W b_+ +(\chi W +U V \cos\theta_H)b_-]
           +{W \over s_H} f \biggr\} & \eqnlabel{G3} \cr
 G_4 = & U s_H \sqrt{s_L}
         \biggl(a_- + {m_B{}^2 +s_H -s_L \over 2 s_H}b_- \biggr) &
         \eqnlabel{G4} \cr
 G_5 = & {1 \over \sqrt{s_H}}
         [W^2 b_+ + W(\chi W +U V \cos\theta_H)b_- + V f] &
         \eqnlabel{G5} \cr
 G_6 = & U W \sqrt{s_L} b_- & \eqnlabel{G6} \cr
 G_7 = & \sqrt{s_L} f  & \eqnlabel{G7} \cr
 G_8 = & {U W \sqrt{s_H s_L} \over 2 m_{D^*}}
         \biggl[g_+ - g_- + (\lambda s_H)u_1 \cr
          &\msp{6} +\biggl({m_B{}^2 +s_H -s_L \over 2}\lambda
            + U W \cos\theta_H \biggr)u_2 \biggr]  & \eqnlabel{G8} \cr
 G_9 = & \sqrt{s_L} [W g_+ + (\chi W + U V \cos\theta_H)g_-
                     + (U s_H \cos\theta_H)r] & \eqnlabel{G9} \cr
 G_{10} = & U \sqrt{s_H} (s_L g_- + V r) & \eqnlabel{G10} \cr
 G_{11} = & U W \sqrt{s_L}
          \biggl[g_- - \biggl(s_H u_1
                 + {m_B{}^2 +s_H -s_L \over 2}u_2 \biggr)\biggr]
          & \eqnlabel{G11} \cr
 G_{12} = &  U V \sqrt{s_L} (g_- - V u_2) & \eqnlabel{G12} \cr
 G_{13} = &  U \sqrt{s_H} s_L (g_- - V u_2) & \eqnlabel{G13} \cr
 G_{14} = &  U V \sqrt{s_H} (r + s_L u_2) & \eqnlabel{G14} \cr
 G_{15} = &  U s_H \sqrt{s_L} (r + s_L u_2) & \eqnlabel{G15} \cr
 G_{16} = &  U W \sqrt{s_H} [W w_+ + (\chi W +U V \cos\theta_H)w_-] &
             \eqnlabel{G16} \cr
 G_{17} = &  U^2 W s_H \sqrt{s_L} w_- & \eqnlabel{G17} \cr}$$
\end{eqnseries}
In these equations, $\lambda = 1 + \chi$.

\newpage
	Then
\begin{eqnseries}[Ij]
$$\eqalignno{
 I_1 = & {1 \over 2} (|G_1|^2 - |G_3|^2 + |G_5|^2)
         + {3 \over 2}|G_7|^2 + {3 \over 4}|G_9|^2 \cr
       & +{3 \over 4}(|G_2|^2 - |G_4|^2 + |G_6|^2 + |G_8|^2
          + |G_{10}|^2 \cr
       & \msp{2} - |G_{11}|^2 + |G_{12}|^2 - |G_{13}|^2
          - |G_{14}|^2 + |G_{15}|^2) \sin^2\theta_H \cr
       & +{1 \over 2}|G_{10} + G_{16}|^2 \sin^2\theta_H
         +{3 \over 4}|G_9 - G_{17} \sin^2\theta_H|^2 &
      \eqnlabel{I1} \cr
 I_2 = & -{1 \over 2} (|G_1|^2 - |G_3|^2 + |G_5|^2)
         + {1 \over 2}|G_7|^2 + {1 \over 4}|G_9|^2 \cr
       & +{1 \over 4}(|G_2|^2 - |G_4|^2 + |G_6|^2 + |G_8|^2
         + |G_{10}|^2 \cr
       & \msp{2} - |G_{11}|^2 + |G_{12}|^2 - |G_{13}|^2
          - |G_{14}|^2 + |G_{15}|^2) \sin^2\theta_H \cr
       & -{1 \over 2} |G_{10} + G_{16}|^2 \sin^2\theta_H
         +{1 \over 4} |G_9 - G_{17} \sin^2\theta_H|^2
       & \eqnlabel{I2} \cr
 I_3 = & {1 \over 2}(-|G_2|^2 + |G_4|^2 - |G_6|^2 + |G_8|^2
          + |G_{10}|^2 \cr
       & \msp{1} - |G_{11}|^2 + |G_{12}|^2 - |G_{13}|^2
          - |G_{14}|^2 + |G_{15}|^2) \sin^2\theta_H \cr
       & -{1 \over 2} |G_{17}|^2 \sin^4\theta_H & \eqnlabel{I3} \cr
 I_4 = & {\rm Re}(G_1 G_2^* - G_3 G_4^* + G_5 G_6^* - G_9 G_{10}^*)
         \sin\theta_H \cr
         &+{\rm Re}(G_{16} G_{17}^*) \sin^3\theta_H & \eqnlabel{I4} \cr
 I_5 = & 2{\rm Re}[G_1 G_8^* + G_3 G_{11}^* - G_5 (G_{12} + G_{15})^*
         - G_7 (G_{10} + G_{16})^*] \sin\theta_H & \eqnlabel{I5} \cr
 I_6 = & 2 {\rm Re}\{[G_2 G_8^* + G_4 G_{11}^* - G_6 (G_{12} + G_{15})^*]
                \sin^2\theta_H + 2 G_7 G_9^* \cr
                &\msp{3}- G_7 G_{17}^* \sin^2\theta_H \} &\eqnlabel{I6}\cr
 I_7 = & 2 {\rm Im}(G_1 G_2^* - G_3 G_4^* + G_5 G_6^* + G_9 G_{10}^*)
         \sin\theta_H & \eqnlabel{I7} \cr
 I_8 = & {\rm Im}[G_1 G_8^* + G_3 G_{11}^* - G_5 (G_{12} + G_{15})^*
             +G_7 (G_{13} + G_{14})^*] \sin\theta_H
       & \eqnlabel{I8} \cr
 I_9 = & - {\rm Im} [G_2 G_8^* + G_4 G_{11}^* - G_6 (G_{12} + G_{15})^*]
         \sin^2\theta_H & \eqnlabel{I9} \cr}$$
\end{eqnseries}

	Eqs.(\puteqn{Ij}) indicates that the partial wave expansions
for the $G_i$ in eqs.(\puteqn{G}) are of the form
\begin{eqnseries}[Gexpn]
$$\eqalignno{
 G_i(s_H,s_L,\cthH) = & \sum_{l=0}^{\infty} \Gt{i}(s_H,s_L)
                      P_l(\cthH), & \eqnlabel{G1expn} \cr
   &\hbox{for $i=1,3,5,7,9,$} \cr
 G_i(s_H,s_L,\cthH) = & \sum_{l=1}^{\infty}
                      {\Gt{i}(s_H,s_L) \over \sqrt{l(l+1)}}
                      {d \over d \cthH}
                      P_l(\cthH), & \eqnlabel{G2expn} \cr
   &\hbox{for $i=2,4,6,8,10,11,12,13,14,15,16,$ and} \cr
 G_{17}(s_H,s_L,\cthH) = & \sum_{l=1}^{\infty}
                      {\Gt{17}(s_H,s_L) \over
                       \sqrt{(l-1)l(1+1)(l+2)}} {d^2 \over d^2 \cthH}
                      P_l(\cthH). & \eqnlabel{G17expn} \cr}$$
\end{eqnseries}

	The form of the distribution given by
eq.(\puteqn{d5Gamma}-\puteqn{Gexpn}), where the dependence on the
lepton angles ($\theta_L,\phi$) is explicit, is ideally suited for the
determination of the $I_j$'s and hence the form factors from  angular
distribution measurements.

	Implementing eqs.(\puteqn{I},\puteqn{Ij},\puteqn{Gexpn}) in
\puteq{d5Gamma} and integrating over the angles yields
$$\eqalignno{
 d^2 \Gamma = & {G_F{}^2 |V_{cb}|^2 \over 3(4 \pi)^5 m_B^3} U W\cr
  &\sum_l {2 \over 2l+1}
   \Bigl[|\Gt{1}|^2 -|\Gt{3}|^2 +|\Gt{5}|^2 +2|\Gt{7}|^2 +|\Gt{9}|^2
         +|\Gt{9} -\Gt{17}|^2 \cr
  &\msp{5} +|\Gt{2}|^2 -|\Gt{4}|^2 +|\Gt{6}|^2 +|\Gt{8}|^2
           +|\Gt{10}|^2 -|\Gt{11}|^2 \cr
  &\msp{5} +|\Gt{12}|^2 -|\Gt{13}|^2 -|\Gt{14}|^2 +|\Gt{15}|^2
           +|\Gt{10} +\Gt{16}|^2 \Bigr];
  &\eqnlabel{d2aGamma}\cr}$$
and the total decay rate is
$$\Gamma = \int_{(\mDs +m_X)^2}^{m_B{}^2}
 \biggl[\int_0^{(m_B -\sqrt{s_H})^2}
 \biggl({d^2 \Gamma \over d s_H d s_L}\biggr) d s_L\biggr] d s_H.
 \eq{Gamma}
The simplicity of the limits in the integration over phase space in
\puteq{Gamma} is another advantage of our choice of the five
variables describing four body decay.

\sec{$B \rightarrow D^* X \ell \bar \nu_\ell$ DECAY}

	Weak $b \rightarrow c$ transitions like the decay
$B \rightarrow D^* X \ell \bar \nu_\ell$ are effected by the current
$\bar c \gamma_{\mu} (1 - \gamma_5) b$. Since this operator is
a singlet under $\rm SU(3)_L \times SU(3)_R$ chiral transformations,
matching its matrix elements onto those in the heavy meson
chiral theory gives \ref{Wis92}
$$\bar c \gamma_{\mu} (1 - \gamma_5) b =
 -C_{cb}\beta(v \cdot v^\prime) {\rm Tr}[\bar H_a^{(c)}(v^\prime)
\gamma_{\mu} (1 - \gamma_5) H_a^{(b)}(v)] + \ldots, \eq{VAcurr}
where the ellipsis denotes terms with derivatives, insertions
of the light quark mass matrix $m_q$, or factors of $1/m_Q$.
The factor $C_{cb}$ contains the calculable perturbative QCD
corrections to the heavy quark current, while $\beta$ is the
universal Isgur-Wise meson form factor which accounts for the
unknown non-perturbative effects in the current due to interactions
with the
light degrees of freedom. The function $\beta$ is normalized at
zero recoil:\ref{Isg89&90}
$$\beta(v \cdot v^\prime = 1) = 1 \eq{betanorm}

	At leading order in the heavy meson chiral perturbation
theory, $\pi,K,\eta$ fields are absent in the operator of
\puteq{VAcurr}, and hence the matrix element for
$B \rightarrow D^* X \ell \bar\nu_\ell$ decay is dominated by the
tree-level pole-type Feynman graphs in \Figure{BDXlndiag}. The Feynman
rules for these diagrams are obtained by expanding out \puteq{L_HMCS}
and (\puteqn{VAcurr}) in powers of the pseudo-Goldstone boson fields
and the heavy meson fields $P_a$ and $P_{a\mu}^*$.

	Calculating the Feynman diagrams for the case
$X = \pi^{\pm}$ gives the following predictions for the form
factors.
\begin{eqnseries}[formfact]
$$\eqalignno{
 a_+ &= A \left({1 \over 2 \,\mDs} + {1 \over m_B}\right)
  \left({1 \over \Dprop2} - {1 \over \Bprop}\right) & \eqnlabel{a+} \cr
 a_- &= {A \over 2 \,\mDs}
        \left({1 \over \Dprop2} - {1 \over \Bprop}\right)
   & \eqnlabel{a-} \cr
 b_+ &= {A \over 2 \, m_B}\left[{1 \over \mDs}
  \left(1 + {v \cdot p_{\pi} \over \Bprop} \right)
  +\left({1 \over \Bprop} - {1 \over v^\prime \cdot p_{\pi}} \right)
  \right] & \eqnlabel{b+} \cr
 b_- &= {A \over 2 \, m_B}\left[{1 \over \mDs}
  \left(1 + {v \cdot p_{\pi} \over \Bprop} \right)
  -\left({1 \over \Bprop} - {1 \over v^\prime \cdot p_{\pi}} \right)
  \right] & \eqnlabel{b-} \cr
 c &= {A \over m_B} \left({1 \over \Dprop2} - {1 \over \Bprop} \right)
  & \eqnlabel{c} \cr
 d &= 0 & \eqnlabel{d} \cr
 f &= A \left[{v^\prime \cdot p_\pi
               - (v \cdot v^\prime)(v \cdot p_{\pi}) \over \Bprop}
  + {v \cdot p_\pi \over v^\prime \cdot p_{\pi}} - v \cdot v^\prime \right]
  & \eqnlabel{f} \cr
 g_+ &= -{A \over 2\>m_B} \left[{1 \over \mDs}
  \left({v \cdot p_{\pi} \over \Bprop}\right)
  + {1 \over \Bprop} \right] & \eqnlabel{g+} \cr
 g_- &= -{A \over 2\>m_B} \left[{1 \over \mDs}
  \left({v \cdot p_{\pi} \over \Bprop}\right)
  - {1 \over \Bprop} \right] & \eqnlabel{g-} \cr
 r &= -{A \over 2} \Biggl\{{1 \over m_B} \left[{1 \over \mDs}
  \left({v \cdot p_{\pi} \over \Bprop}\right)
  - {1 \over \Bprop} \right] \cr
  &\msp{3} +{1 \over \mDs} \left({1 + v \cdot v^\prime \over
  v^\prime \cdot p_{\pi}} - {1 \over \Bprop}\right) \Biggr\}
  & \eqnlabel{r} \cr
 u_1 &= 0 & \eqnlabel{u1} \cr
 u_2 &= 0 & \eqnlabel{u2} \cr
 w &= 0 & \eqnlabel{w} \cr
 w_+ &= w_- = {A \over 2 \> m_B \> \mDs^2 \> v^\prime \cdot p_{\pi}}
  & \eqnlabel{w+-} \cr}$$
\end{eqnseries}
In these equations,
\subequationnumstyle{alphabetic}
$$\eqalignno{
 A &= \sqrt{\mDs \> m_B} \>g \>C_{cb} \>\beta(v \cdot v^{\prime})/f_{\pi}
  &\eqnlabel{A}\cr
 \Delta_D &= \mDs - m_D \approx 142 \,\rm MeV, & \eqnlabel{Dsplit} \cr
 \Delta_B &= m_{B^*} - m_B \approx 46 \,\rm MeV. & \eqnlabel{Bsplit}
  \cr}$$
\subequationnumstyle{blank}
Multiplying the above expressions by the factor $\pm 1/\sqrt2$ gives
the corresponding form factors for a neutral pion.

	The above results are generally applicable when $X$ is any
of the pseudo-Goldstone bosons with appropriate modifications to take
into account isospin factors. However, the large masses of the kaon and
eta compared to the chiral symmetry breaking scale
($\Lambda_{\chi} \sim 1 \rm\> GeV$) may render
leading order chiral perturbation theory inadequate, so
in the remainder of this analysis we will continue to take $X$
to be a pion.

	Since the masses of the heavy mesons are so much greater
than that of the pseudo-Goldstone bosons, it is appropriate to make
the dependence on the heavy masses manifest and to neglect terms
that are suppressed by factors of $m_{\pi}/m_B$ and $m_{\pi}/\mDs$.
The pertinent formulae in Sect.2 can be written in this form by
expressing the pion's four-momentum in terms of its four-velocity
$v^\mu_{\pi} = p^\mu_{\pi}/m_{\pi}$ and by changing variables from
$s_H$ and $s_L$ to $v \cdot v_{\pi}$ and $v^{\prime} \cdot v_{\pi}$
so that the integration measure in \puteq{d5Gamma} becomes
$$d s_H \> d s_L \approx 4 \> m_B \> \mDs^2 \> m_{\pi}
  d(v \cdot v^\prime) \> d(v^{\prime} \cdot v_{\pi}). \eq{dsHsL}
Now we introduce the dimensionless quantities $\hat G_j$ which are
defined in terms of the $G_j$ by
$$G_j = {m_B^{3/2} \> \mDs^{1/2} \> g \> C_{cb} \>
   \beta(v \cdot v^{\prime}) \over f_{\pi}} \Gh_j \eq{Ghatj}
into \puteq{d5Gamma}. Substituting
$$ U \approx {2 \, m_{\pi} \over \mDs}
       [(v^{\prime} \cdot v_{\pi})^2 - 1]^{1/2},$$
$$ W \approx m_B \> \mDs [(v \cdot v^{\prime})^2 - 1]^{1/2},$$
and performing the integrations over $\theta_L$ and $\phi$ in the
differential decay rate in \puteq{d5Gamma} yields
$$\eqalignno{
 d^3 \Gamma = & {8 G_F^2 m_B^2 \mDs^3 |V_{cb}|^2 \over 3(4\pi)^5}
  \left({m_{\pi} \over f_{\pi}}\right)^2 g^2 C_{cb}{}^2
  \beta(v \cdot v^{\prime})^2 [(v \cdot v^{\prime})^2 - 1]^{1/2}
  [(v^\prime \cdot v_\pi)^2 - 1]^{1/2} \cr
 &\big[\big(|\Gh_1|^2 -|\Gh_3|^2 +|\Gh_5|^2 +2|\Gh_7|^2 +|\Gh_9|^2
    +|\Gh_9 - \Gh_{17}\sin^2 \theta_H|^2 \big) \cr
   &+\big(|\Gh_2|^2 -|\Gh_4|^2 +|\Gh_6|^2 +|\Gh_8|^2 +|\Gh_{10}|^2 \cr
     &\msp{1}-|\Gh_{11}|^2 +|\Gh_{12}|^2 -|\Gh_{13}|^2 -|\Gh_{14}|^2 \cr
     &\msp{1}+|\Gh_{15}|^2 +|\Gh_{10} +\Gh_{16}|^2 \big) \sin^2 \theta_H
  \big]\>
 d(v \cdot v^{\prime}) \, d(v^{\prime} \cdot v_{\pi}) \, d \cos\theta_H,
 &\eqnlabel{d3Gamma}\cr}$$
where
$$v \cdot v_{\pi} = (v \cdot v^{\prime})(v^{\prime} \cdot v_{\pi})
  - [(v \cdot v^{\prime})^2 -1]^{1/2}
    [(v^{\prime} \cdot v_{\pi})^2 -1]^{1/2} \cos\theta_H. \eq{vvpi}

	A source of uncertainty in \puteq{d3Gamma} is the
Isgur-Wise function $\beta(v \cdot v^{\prime})$ since its value is
only known at the zero recoil point given by \puteq{betanorm}. However,
the quantity $v \cdot v^{\prime}$ is unconstrained, so
this dependence on $\beta$ can be removed by normalizing this decay
rate to that for the corresponding semi-leptonic transition
without the emission of pseudo-Goldstone bosons:
$$B \rightarrow D^* \ell \bar\nu_{\ell} \eq{BDsln}
This transition is mediated by the current in \puteq{VAcurr} and
the hadronic matrix element is
$$\eqalignno{
 &\bra{D^*(v^\prime,\epsilon)} \bar c \gamma_{\mu} (1 - \gamma_5) b
 \ket{B(v)} \cr
 =& \sqrt{m_B m_{D^*}} \> C_{cb} \> \beta(v \cdot v^\prime) \>
  [-(1 + v \cdot v^\prime)\epsilon_{\mu}^*
  + (\epsilon^* \cdot v) v_{\mu}^\prime
  + i \epsilon_{\mu\alpha\beta\gamma} \epsilon^{*\alpha}
      v^{\prime \beta} v^{\gamma}]. &\eqnlabel{BDln}\cr}$$
Then the rate could be studied away from the zero recoil point.

	Since the above rate
involves the ratio $m_{\pi}/f_{\pi}$ which is close to unity, and
is not suppressed by heavy quark masses, the rate for the decay
as given by \puteq{d3Gamma} is appreciable in the region of phase
space where chiral perturbation theory is valid.
To show this, we
introduce a scaled decay rate $d^3\hat\Gamma$ defined by
$$d^3\Gamma = {G_F^2 m_B^5 \over 192 \pi^3}|V_{cb}|^2 g^2 C_{cb}{}^2
 \beta(v \cdot v^{\prime})^2 d^3 \hat\Gamma. \eq{d3hGamma}
The differential rate
$d^2 \hat\Gamma / d(v \cdot v^{\prime}) d(v^{\prime} \cdot v_{\pi})$
is calculated for various values of $v \cdot v^{\prime}$ and
$v^{\prime} \cdot v_{\pi}$ in Table 1. In \Figure{d2GammaPlot},
this rate is plotted as a function of $v \cdot v^\prime$ and
$v^\prime \cdot v_\pi$ in the kinematic region where chiral
perturbation theory is expected to be valid. Additional plots of
the differential decay rate as a function of the various kinematic
variables, which may be of relevance to experimental analyses, are not
presented because in practice such analyses are typically performed by
doing Monte Carlo simulations of the fully differential decay rate
given by \puteq{d5Gamma}.

	Table 1 shows that the differential rate for
$B \rightarrow D^* \pi \ell \bar \nu_\ell$ decay is smaller than the
corresponding rate for $B \rightarrow D \pi \ell \bar \nu_\ell$
decay given in Table 1 of Ref.[\putref{Lee92}]. This enhancement for
$B \rightarrow D \pi \ell \bar \nu_\ell$ can be attributed in part
to the $D^*$ propagator, in Fig.2 of Ref.\putref{Lee92}, which
becomes on-shell as its pole is approached. However, the
presence of the $D^*$ in the decay
$B \rightarrow D^* \pi \ell \bar \nu_\ell$ allows this process to
be selected experimentally with much better signal to background
(because of the small amount of phase space available for
$D^* \rightarrow D \pi$ decay) as compared to the decay mode
$B \rightarrow D \pi \ell \bar \nu_\ell$. Moreover, the decay rate
for the former channel increases much more rapidly with
$v \cdot v^\prime$ than in the latter channel. So an experimental
study of $B \rightarrow D^* \pi \ell \bar \nu_\ell$ decay would
complement a similar study of
$B \rightarrow D \pi \ell \bar \nu_\ell$. A measurement of this
decay rate could be used to test heavy quark flavor symmetry:
if this symmetry were violated, there would be different couplings
$g_c$ and $g_b$ for the $D^*D\pi$ and $B^*B\pi$ vertices in
Fig.\putlab{BDXlndiag} which would result in different
expressions for the form factors in eqs.(\puteqn{formfact})
and hence in a different decay rate.

	The value that the differential decay rate takes is
determined by the contributions coming from the pole-type graphs in
Fig.\putlab{BDXlndiag}. In order for these pole diagrams to be the
dominant contribution to the perturbative chiral expansion,
the pseudo-Goldstone boson must be emitted with low momentum. Or
equivalently, the chiral expansion parameters
$v \cdot p_{\pi}/ \Lambda_{\chi}$ and
$v^{\prime} \cdot p_{\pi}/ \Lambda_{\chi}$
should be small --- with $v \cdot p_\pi$ and $v^\prime \cdot p_\pi$
on the order of a few hundred MeV. An attempt to
estimate the regime where chiral perturbation theory is valid for the
decay $B \rightarrow D \pi \ell \bar \nu_\ell$ was made in
Ref.[\putref{Lee92}]; in this analysis it was found that predictions of
next-to-leading order effects in chiral perturbation theory could not
be made because there were too many higher dimension operators with
unknown coefficients. A similar study here yields the same result, but
the predictions made in this paper on the basis of leading order
chiral perturbation theory may well be valid over a kinematic range
much larger than that exhibited in Table 1. An experiment would
ultimately establish the region of phase space where our results are
valid.

\sec{CONCLUDING REMARKS}
	In this paper, a complete kinematical analysis for
$B \rightarrow D^* X \ell \bar \nu_\ell$ decay is presented. The
constraints that the heavy quark and chiral symmetries impose on
this decay are found to considerably simplify the dynamics and
are used to determine the decay rate for this process. A number of
extensions to this work can be pursued. For instance, it is
interesting to determine how large symmetry-breaking effects are
by calculating sub-leading $\Lambda_{\rm QCD}/m_c$ corrections.
Decays in which more than one pseudo-Goldstone boson is emitted
can also be considered.

\medskip

{\it Note added.}
	After the completion of this work, a short paper by H.-Y. Cheng
[preprint IP-ASTP-18-92] appeared which also considers the decay analyzed
here.

\sec{Acknowledgement}
The author would like to thank Mark Wise for helpful conversations.

% TABLES

\newpage   \sectionnumstyle{blank}
\centerline{\bf Table 1}
\noindent Differential decay rate
$d^2 \hat\Gamma / d(v \cdot v^{\prime}) d(v^{\prime} \cdot v_{\pi})$
at different values of $v \cdot v^{\prime}$ and
$v^{\prime} \cdot v_{\pi}$.
\medskip
\begintable
$v \cdot v^{\prime}$|$v^{\prime} \cdot v_{\pi}$|
$d^2 \hat\Gamma / d(v\cdot v^{\prime}) d(v^{\prime}\cdot v_{\pi})$\crthick
1.2|1.1|0.0022\cr
1.4|1.1|0.0037\cr
1.2|1.2|0.0033\cr
1.4|1.2|0.0057\cr
1.2|1.3|0.0043\cr
1.4|1.3|0.0074\cr
1.1|1.4|0.0022\cr
1.2|1.4|0.0052\cr
1.4|1.4|0.0090\cr
1.2|1.5|0.0061\cr
1.4|1.5|0.011\endtable

% REFERENCES

\newpage  \sectionnumstyle{blank}
\sec{REFERENCES}

\begin{putreferences}
\reference{Wis91}{M.B. Wise, \underbar{New Symmetries of the Strong
                  Interaction}, Lectures presented at the 1991 Lake Louise
                  Winter Institute, CALT-68-1721.}
\reference{Wis92}{M.B. Wise, \journal{Phys. Rev. D}{45}(1992) R2188.}
\reference{Bur92}{G. Burdman and J.F. Donoghue, \journal{Phys. Lett.}
                  {B280}(1992) 287.}
\reference{Yan92}{T.-M. Yan, H.-Y. Cheng, C.-Y. Cheung, G.-L. Lin, Y.C. Lin,
                  and H.-L. Yu, \journal{Phys. Rev. D}{46}(1992) 1148.}
\reference{Cho92}{P. Cho, \journal{Phys. Lett.}{B285}(1992) 145.}
\reference{Lee92}{C.L.Y. Lee, M. Lu and M.B. Wise, {\sl Phys. Rev. D}
                  (in press), CALT-68-1771 (1992).}
\reference{ACC92}{ACCMOR Collaboration, S. Barlag {\it et al.},
                  \journal{Phys. Lett.}{B278}(1992) 480.}
\reference{Cab65}{N. Cabibbo and A. Maksymowicz, \journal{Phys. Rev.}{137}
                  (1965) B438;
                  A. Pais and S.B. Treiman, {\it ibid.} {\bf 168}(1968)
                  1858.}
\reference{Isg89&90}{N. Isgur and M.B. Wise, \journal{Phys. Lett.}{B232}
                     (1989) 113;\\
                     N. Isgur and M.B. Wise, \journal{Phys. Lett.}{B237}
                     (1990) 527.}
\end{putreferences}

\newpage
\subsec{Figure Captions}

\item{Figure \putlab{BDXlndiag}.}
 Leading order Feynman diagrams for
 $B \rightarrow D^* X \ell \bar\nu_\ell$ decay.
 The shaded circle represents an interaction term coming from the heavy
 meson chiral Lagrangian of \puteq{L_HMCS}, and the shaded box
 denotes an insertion of the weak current given by \puteq{VAcurr}.
\medskip
\item{Figure \putlab{d2GammaPlot}.}
 The scaled differential decay rate
 $d^2 \hat\Gamma / d(v \cdot v^\prime) d(v^\prime \cdot v_\pi)$
 for $B \rightarrow D^* \pi \ell \bar\nu_\ell$ decay
 plotted as a function of $v \cdot v^\prime$ and $v^\prime \cdot v_\pi$.

\bye